\documentclass[%
 reprint,
 amsmath,amssymb,
 aps,
]{revtex4-2}

\usepackage{graphicx}% Include figure files
\usepackage{dcolumn}% Align table columns on decimal point
\usepackage{bm}% bold math
\usepackage{amsmath, amsthm, amssymb, amsfonts,mathrsfs,mathtools}

\usepackage[linktocpage=true]{hyperref}
\hypersetup{
	colorlinks,
	citecolor=blue,
	filecolor=black,
	linkcolor=blue,
	urlcolor=blue
}
\usepackage{url}
  
\usepackage{physics}
\usepackage{nicematrix}
\usepackage{soul}
\usepackage{array}          
\usepackage{scalerel,stackengine}
\newcolumntype{C}[1]{>{\centering\arraybackslash$}m{#1}<{$}}
\newlength{\mycolwdgb}                                         % array column width
\newcommand\pig[1]{\scalerel*[5pt]{\big#1}{%
  \ensurestackMath{\addstackgap[1.5pt]{\big#1}}}}

\newcommand{\quads}[1][1]{\hspace*{#1em}\ignorespaces}

\begin{document}

\preprint{APS/123-QED}

%\title{Interference phenomena in the asymmetric dynamical Casimir effect\\ for a single $\ddp$ mirror}

\title{On the Lorentz Boosted Parallel Plate Casimir Cavity}

\author{Matthew~J.~Gorban}
\email{matthew\_gorban1@baylor.edu}%
\author{William~D.~Julius}
\email{william\_julius1@baylor.edu}%
\author{Gerald~B.~Cleaver}
\email{gerald\_cleaver@baylor.edu}%
\affiliation{%
 Department of Physics, Baylor University,  Waco, TX 76798, USA 
}%

\date{\today}% It is always \today, today,
             %  but any date may be explicitly specified

\begin{abstract}
Two perfectly conducting, infinite parallel plates will restrict the electromagnetic vacuum, producing an attractive force. This phenomenon is known as the Casimir effect. Here we use electromagnetic field correlators to define the local interaction between the plates and the vacuum, which gives rise to a renormalized stress-energy tensor. We then show that a Lorentz boost of the underlying electric and magnetic fields that comprise the correlators will produce the correct stress-energy tensor in the boosted frame. The infinite surface divergences of the field correlators will transform appropriately, such that they cancel out in the boosted frame and produce the desired finite result.
\end{abstract}

%\keywords{Suggested keywords}%Use showkeys class option if keyword
                              %display desired
\maketitle

%\tableofcontents

\textit{Introduction}.---First introduced by Hendrick Casimir in 1948 \cite{casimir1948attraction}, the boundaries of an enclosed cavity in free-space will impose strict limitations on the underlying quantum vacuum and restrict fundamental vacuum modes of the background vacuum field. For the simplest cavity configuration, composed of two infinite, perfectly conducting parallel plates located a distance $a$ from each other, there is a resulting force per unit area of
\begin{equation*}
    \frac{F}{A}=-\frac{\pi^2\hbar c}{240 a^4}.
\end{equation*}
This force, duly named the Casimir force, seeks to push the plates together (evident by the minus sign) and, remarkably, the interior of the plates has a corresponding \textit{negative} energy density. This phenomenon, the Casimir effect, is not limited to just simple cavity configurations. It extends to a wide variety of physical interactions between the quantum vacuum and surfaces with various geometries and boundary conditions (or physically, the properties of the materials constituting that surface). The Casimir effect is commonly referred to as a physical manifestation of the quantum vacuum \cite{milton2001casimir,milonni1992casimir,milton2004casimir,lamoreaux2004casimir,bordag2009advances,simpson2015forces,palasantzas2015casimir}. There is an abundance of literature that encompasses the extensive influence of the Casimir effect to many aspects of physics: see \cite{mostepanenko1997casimir,milton2004casimir,bordag2009advances,dalvit2011casimir} for detailed reviews of this topic. 

The Casimir effect is not limited to just theoretical perspectives; there is an extensive history of experimentation that ranges from advancing force measurements \cite{sparnaay1958measurements,lamoreaux1997demonstration,mohideen1998precision,chan2001quantum} to the consequences of dielectric response \cite{decca2003measurement,iannuzzi2004effect,munday2009measured,banishev2013measuring,somers2018measurement} to the effects of geometry \cite{bressi2002measurement,chan2008measurement,tang2017measurement,garrett2018measurement}. It has even been shown that the Casimir effect can facilitate heat transfer across the vacuum via quantum fluctuations between nanomechanical systems \cite{fong2019phonon}. In fact, Casimir experiments are no longer limited to the static Casimir effect. When a mirror interacting with the vacuum is subjected to time-dependent boundary conditions, the system produces real photons by means of the dynamical Casimir effect \cite{dodonov2009dynamical,dodonov2010current,dodonov2020fifty}. In 2011, the first experimental detection of this effect was performed. In this experiment, photon production was observed by utilizing a superconducting circuit whose electrical length can be changed by modulating the inductance of
a superconducting quantum interference device (SQUID) at high frequencies \cite{wilson2011observation}. For a modern review of Casimir experimentation, see \cite{gong2020recent}.

When computing the electromagnetic Casimir effect (specifically the vacuum expectation values of the Maxwell stress tensor or the relativistic stress-energy tensor) it is important to correctly account for the local behavior due to the presence of strong divergences as we approach the boundaries \cite{deutsch1979boundary,santos2005electromagnetic}. It is possible to remove these infinities if one chooses a suitable method of locally defining the field, seen in Brown and Maclay's original derivation of the stress-energy tensor for the parallel plate cavity \cite{brown1969vacuum}. However, the electromagnetic field correlators we will use to define the local interaction between the field and the plates contain divergences that cannot be removed with standard renormalization methods \cite{santos2005electromagnetic}. These correlators encode the entire local behavior near both sides of the material boundaries and can be used to directly calculate the components of the stress-energy of the Casimir cavity.

In this Letter, we show that one can apply a local Lorentz transformation (boost) to the underlying electric and magnetic fields that compose the electromagnetic field correlators of a parallel plate Casimir cavity and recover the resulting transformation applied directly to the stress-energy tensor. From Brown and Maclay's derivation \cite{brown1969vacuum}, the renormalized stress-energy tensor for perfectly conducting parallel plates, separated by distance $a$, takes the form
\begin{equation}\label{RSET}
    \expval*{\hat{\Theta}^{\mu\nu}(z)}^{ren}_0=\frac{\pi^2\hbar c}{720a^4}\mathrm{diag}(-1,1,1,-3).
\end{equation}
This stress-energy tensor will transform according to the standard methods employed in special relativity when the Lorentz boost is applied in the perpendicular direction relative to the surface of the plates. It should be expected that applying a Lorentz boost to the electric and magnetic field correlators will produce the same result. Thus, we present an explicit derivation that shows this is indeed the case and that the divergent infinities present in the field correlators also transform correctly, such that they exactly cancel each other out and the resulting stress-energy tensor in the boosted frame remains a finite result. While this result is not surprising, the authors are not aware of any existing literature that explicitly demonstrates the relativistic transformation of Casimir setups. 

Here we will be using the metric signature convention $(-1,1,1,1)$. Einstein summation notation will be used, where Greek indices will run from 0 to 3 and Latin indices will run from 1 to 3, with the exception of $\alpha$ which is used to track the modal function of the electromagnetic field. Gaussian units are employed throughout and we will take $c=\hbar=1$ hereafter.

%%%%%%%%%%%%%%%%%%%%%%%%%%%%%
%%%%%%%%%%%%%%%%%%%%%%%%%%%%%

\textit{Parallel Plate Casimir Cavities}.---The classical Maxwell stress tensor for any system is given by \cite{jackson1999classical}
\begin{equation}\label{Maxwell classical}
    T_{ij}=\frac{1}{4\pi}\qty(E_iE_j-\frac{1}{2}\delta_{ij}\vectorbold{E}^2+B_iB_j-\frac{1}{2}\delta_{ij}\vectorbold{B}^2),
\end{equation}
where $\vectorbold{E}$ and $\vectorbold{B}$ are the electric and magnetic fields, denoted in bold to indicate vector quantities. Our first task is to express this quantity in terms of the Casimir electromagnetic field correlators. To facilitate computation we will assume that the parallel plate's cavity walls will be oriented such that the surfaces lie parallel to the $xy$-plane, and perpendicular to the $z$-axis.

The quantum version of the stress tensor in Eq. \eqref{Maxwell classical} can be obtained by quantizing the electromagnetic field \cite{santos2005electromagnetic}, which becomes 
\begin{equation}\label{Maxwell quantum}
    4\pi\expval*{\hat{T}_{ij}}_0=\expval*{\hat{E}_{i}\hat{E}_{j}}_0-\frac{1}{2}\delta_{ij}\expval*{\vectorbold{\hat{E}}{}^2}_0+\expval*{\hat{B}_{i}\hat{B}_{j}}_0-\frac{1}{2}\delta_{ij}\expval*{\vectorbold{\hat{B}}{}^2}_0.
\end{equation}
Here we employ the shorthand notation $\expval*{\hat{A}}_0\equiv\expval{\hat{A}}{0}$ to denote the vacuum expectation value of some operator $\hat{A}$ throughout this paper. We will use this version of the Maxwell tensor to find the spatial components of the renormalized stress-energy tensor. For example,
\label{sec:headings}
\begin{equation}
    \expval*{\hat{T}_{zz}}_0= \frac{1}{8 \pi}\bqty{\expval*{\hat{E}^2_{z}}_0-\expval*{\hat{E}^2_{\parallel}}_0+\expval*{\hat{B}^2_{z}}_0-\expval*{\hat{B}^2_{\parallel}}_0}, 
\end{equation}
where $\hat{E}^2_{\parallel}=\hat{E}^2_{x}+\hat{E}^2_{y}$ and $\hat{B}^2_{\parallel}=\hat{B}^2_{x}+\hat{B}^2_{y}$. 

To find the renormalized stress-energy tensor, $\expval*{\hat{\Theta}^{\mu\nu}(z)}^{ren}_0$, of the stationary Casimir cavity, we will use the relation $\Theta^{ij}(z)=-T_{ij}(z)$ \cite{jackson1999classical} to find the spatial components, along with the energy density, $\rho(z)=\expval*{\hat{\Theta}^{00}(z)}^{ren}_0$, given by
\begin{equation}\label{energy density eq}
    \rho(\vectorbold{r},t)=\frac{1}{8\pi}\qty(\expval*{\vectorbold{\hat{E}}{}^2(\vectorbold{r},t)}_0+\expval*{\vectorbold{\hat{B}}{}^2(\vectorbold{r},t)}_0).
\end{equation}
Crucially, the local nature of the correlators possess strongly divergent behavior near the boundaries of the plates that cannot be renormalized in the usual manner. Yet, the stress-energy tensor will itself be renormalized to a finite value as the divergences resulting from the correlators always appear in pairs which exactly cancel.

In Casimir's original experimental setup \cite{casimir1948attraction}, and in subsequent results from Brown and Maclay \cite{brown1969vacuum}, two infinite perfectly conducting parallel plates are kept at a fixed distance $a$ from each other. We will set this up such that one of the plates is placed at $z=0$ and the other one at $z=a$. The perfectly conducting plates will impose the following boundary conditions on electric and magnetic fields: the tangential components of the electric field, $\hat{E}_x$ and $\hat{E}_y$, along with the normal component of the magnetic field, $\hat{B}_z$, will vanish on the surface of the plates.

A detailed calculation using the cavity setup introduced above, presented by Santos, Sobrinho, and Tort in \cite{santos2005electromagnetic}, yields the following results for the electric and magnetic field correlators:
\begin{eqnarray}
    &&\expval{E_i(\vectorbold{r},t)E_j(\vectorbold{r},t)}_0=\nonumber\\&&\quads[4]\qty(\frac{\pi}{a})^4\frac{2}{3\pi}\qty[\frac{1}{120}\qty(-\delta^{\parallel}_{ij}+\delta_{zij})+\delta_{ij}F(\xi
    )],
\end{eqnarray}
\begin{eqnarray}
    &&\expval{B_i(\vectorbold{r},t)B_j(\vectorbold{r},t)}_0=\nonumber\\&&\quads[4]\qty(\frac{\pi}{a})^4\frac{2}{3\pi}\qty[\frac{1}{120}\qty(-\delta^{\parallel}_{ij}+\delta_{zij})-\delta_{ij}F(\xi)],
\end{eqnarray}
where $\delta^{\parallel}_{ij}\equiv\delta_{xij}+\delta_{yij}$. The function $F(\xi)$ is defined by
\begin{equation}
    F(\xi)\equiv-\frac{1}{16}\frac{d^3}{d\xi^3}\cot(\xi),
\end{equation}
where $\xi\equiv\pi z/a$.  This function is divergent on the bounding plates, i.e. for $\xi\rightarrow0$ and $\xi\rightarrow \pi$, which correspond to $z\rightarrow0$ and $z\rightarrow a$, respectively. These divergences appear such that the divergent pieces cancel away when computing the stress-energy tensor. It is worth while noting that this type of divergence is not directly related to the usual sort of loop divergences appearing in quantum field theory, but is instead related to the idealized nature of our boundary conditions (perfect conductors) which cannot be physically realized \cite{schwinger1978casimir}.

The non-vanishing components of the electric field correlators are
\begin{gather}
    \expval*{\hat{E}^{2}_x(z,t)}_0=\expval*{\hat{E}^{2}_y(z,t)}_0=\qty(\frac{\pi}{a})^4\frac{2}{3\pi}\qty[-\frac{1}{120}+F(\xi)], \nonumber\\
    \expval*{\hat{E}^{2}_z(z,t)}_0=\qty(\frac{\pi}{a})^4\frac{2}{3\pi}\qty[\frac{1}{120}+F(\xi)].
\end{gather}
\noindent Additionally, the non-vanishing components of the magnetic field components are
\begin{gather}
    \expval*{\hat{B}^{2}_x(z,t)}_0=\expval*{\hat{B}^{2}_y(z,t)}_0=\qty(\frac{\pi}{a})^4\frac{2}{3\pi}\qty[-\frac{1}{120}-F(\xi)],\nonumber\\
    \expval*{\hat{B}^{2}_z(z,t)}_0=\qty(\frac{\pi}{a})^4\frac{2}{3\pi}\qty[\frac{1}{120}-F(\xi)].
\end{gather}
\noindent A straightforward calculation leads to the following non-vanishing Maxwell tensor components,
\begin{equation}\label{OG T components}
    \expval*{\hat{T}_{zz}}_0=\frac{\pi^2}{240a^4},\quad\expval*{\hat{T}_{xx}}_0=\expval*{\hat{T}_{yy}}_0=-\frac{\pi^2}{720a^4}.
\end{equation}
These components, along with the energy density
\begin{equation}\label{energy density value}
    \rho(a)=-\frac{\pi^2}{720a^4},
\end{equation}
lead to the well known result for the renormalized stress-energy tensor presented in Eq. \eqref{RSET}.

One can then apply a Lorentz boost to see how $\expval*{\hat{\Theta}^{\mu\nu}(z)}^{ren}_0$ transforms when boosted to a moving frame. To do this, we explicitly transform Eq. \eqref{RSET} by the Lorentz transformation corresponding to a boost following the covariant transformation rule
\begin{equation}\label{LT}
    \Theta^{\,\prime\,\mu\nu}={\Lambda^\mu}_\rho{\Lambda^\nu}_\delta\,\Theta^{\,\rho\delta}.
\end{equation}
In what follows we will take $\boldsymbol{\beta}$ to be the dimensionless boost parameter related to the velocity in our units as $\boldsymbol{\beta}=\boldsymbol{v}$ and $\gamma^{-1}=\sqrt{(1-\boldsymbol{\beta}^2)}$ to be the Lorentz factor.

It is straightforward to verify that Eq. \eqref{RSET} remains unchanged for arbitrary boosts in the $\hat{x}$ and $\hat{y}$ directions, due to the translational invariance along the face of the infinite parallel plates. To illustrate this, consider a boost in the $x$-direction only, noting that any other boost in the $xy$-plane may be obtained by applying a spatial rotation about $z$ prior to the boost. In this case  $\boldsymbol{\beta}=\beta\hat{x}$ with the explicit (passive) boost transformation
\begin{equation}
{{\Lambda}^\mu}_\nu=\qty(\,
\begin{array}{*{4}{@{}C{30pt}@{}}}
\gamma & -\gamma\beta & 0 & 0\\
-\gamma\beta & \gamma &0 & 0 \\
0 & 0 &  1 & 0 \\
0 & 0 & 0 & 1
\end{array}\,).
\end{equation}
Acting this on Eq. \eqref{RSET} gives $\expval*{\hat{\Theta}^{\,\prime\,\mu\nu}(z)}^{ren}_0=\expval*{\hat{\Theta}^{\mu\nu}(z)}^{ren}_0$.

Considering now the somewhat more interesting case of a boost along the $\hat{z}$-direction we have $\boldsymbol{\beta}=\beta\hat{z}$ with the (passive) transformation now taking the form
\begin{equation}\label{zboost}
{\Lambda^\mu}_\nu=\qty(\,
\begin{array}{*{4}{@{}C{\mycolwdgb}@{}}}
\gamma & 0 & 0 & -\gamma\beta\\
0 & 1 & 0 & 0 \\
0 & 0 & 1 & 0 \\
-\gamma\beta & 0 & 0 & \gamma
\end{array}\,).
\end{equation}
Acting this new transformation on Eq. \eqref{RSET} produces the appropriate renormalized stress-energy tensor in the boosted frame,
\begin{eqnarray}\label{Boosted Cavity}
 &&\expval*{\hat{\Theta}^{\,\prime\,\mu\nu}(z)}^{ren}_0=\\[5pt]&&\qquad\frac{\pi^2}{720a^4}
 \begingroup
\setlength\arraycolsep{8pt}
\begin{pmatrix}
-\gamma^2(1+3\beta^2) & 0 & 0 & 4\gamma^2\beta\\
0 & 1 & 0 & 0 \\
0 & 0 & 1 & 0 \\
4\gamma^2\beta & 0 & 0 & -\gamma^2(3+\beta^2)
\end{pmatrix}
\endgroup\nonumber
\end{eqnarray}

%%%%%%%%%%%%%%%%%%%%%%%%%%%%%
%%%%%%%%%%%%%%%%%%%%%%%%%%%%%

\textit{Correlator boost}.---We now seek to show that a transformation of the underlying electric and magnetic fields that compose the correlators will correspond to the standard transformation applied directly to the stress-energy tensor. 

The general (passive) Lorentz transform of $\vectorbold{E}$ is (see 13.3 in \cite{wilcox2016macroscopic})
\begin{equation}\label{E'}
    \vectorbold{E}\xmapsto{\,\,\,\,\Lambda\,\,\,\,}\vectorbold{E^{\,\prime}}=\gamma\qty(\vectorbold{E}+\boldsymbol{\beta}\crossproduct\vectorbold{B})-\frac{\gamma^2}{\gamma+1}\boldsymbol{\beta}\qty(\boldsymbol{\beta}\dotproduct\vectorbold{E}),
\end{equation}
where elements of the boosted frame are denoted by primes. This will transform the electric field correlator in the following way
\begin{eqnarray}
    &&\expval*{\hat{E}^{\,\prime}_i(\vectorbold{r},t) \hat{E}^{\,\prime}_j(\vectorbold{r},t)}_0=\sum_{\alpha} E^{\,\prime}_{i\alpha}(\vectorbold{r})E^{\,\prime*}_{j\alpha}(\vectorbold{r})=\nonumber\\
&&{}\sum_{\alpha}\Big[\gamma^2\Big( E_{i\alpha}E^{*}_{j\alpha}+\qty(\boldsymbol{\beta} \cross \vectorbold{B})_{i\alpha}\qty(\boldsymbol{\beta} \crossproduct \vectorbold{B}^*)_{j\alpha}\nonumber\\
&&\quad\quad\quad\quad{}+ E_{i\alpha}(\boldsymbol{\beta} \cross \vectorbold{B}^*)_{j\alpha}+\qty(\boldsymbol{\beta} \cross \vectorbold{B})_{i\alpha}E^{*}_{j\alpha}\Big)\\
&&\quad\quad{}-{}\frac{\gamma^3}{\gamma+ 1}\Big(\big(E_{i\alpha}\beta_{j}+\qty(\boldsymbol{\beta} \cross \vectorbold{B})_{i\alpha}\beta_{j}\big)\qty(\boldsymbol{\beta} \dotproduct \vectorbold{E}^*_{\alpha})\nonumber\\
&&\quad\quad\quad\quad\quad\quad{}+\qty(\boldsymbol{\beta}\dotproduct \vectorbold{E}_\alpha)\big(\beta_{i}E^*_{j\alpha}+\beta_{i}\qty(\boldsymbol{\beta} \cross \vectorbold{B}^*)_{j\alpha}\big)\Big)\nonumber\\
&&\quad\quad{}+{}\frac{\gamma^4}{\qty(\gamma+ 1)^2}\beta_i\beta_j\qty(\boldsymbol{\beta} \dotproduct \vectorbold{E}_{\alpha})\qty(\boldsymbol{\beta} \dotproduct \vectorbold{E}^*_{\alpha})\Big],\nonumber
\end{eqnarray}

\noindent where the summation on $\alpha$ is over the two modal functions of the fields. The functional dependence on $\vectorbold{r}$ for the electric and magnetic field terms is suppressed within the calculations for compactness. 

Aligning the boost along the $\hat{z}$-direction ($\beta_i=\beta\delta_{iz}$), this correlator becomes
\begin{eqnarray}\label{general E}
&&\expval*{\hat{E}^{\,\prime}_i(\vectorbold{r},t) \hat{E}^{\,\prime}_j(\vectorbold{r},t)}_0=\nonumber\\[5pt]
&&\sum_{\alpha}\Big[\gamma^2\Big( E_{i\alpha}E^{*}_{j\alpha}+\beta\pig(\beta\qty(\hat{z} \cross \vectorbold{B})_{i\alpha}\qty(\hat{z} \crossproduct \vectorbold{B}^*)_{j\alpha}\nonumber\\
&&\quads[4]{}+ E_{i\alpha}(\hat{z} \cross \vectorbold{B}^*)_{j\alpha}+\qty(\hat{z} \cross \vectorbold{B})_{i\alpha}E^{*}_{j\alpha}\pig)\Big)\\
&&\quads[2]{}-\frac{\gamma^3}{\gamma+ 1}\beta^2\Big(\big(E_{i\alpha}\delta_{jz}+\beta\qty(\hat{z} \cross \vectorbold{B})_{i\alpha}\delta_{jz}\big) E^*_{z\alpha}\nonumber\\
&&\quads[7]{}+ E_{z\alpha}\big(\delta_{iz}E^*_{j\alpha}+\beta\delta_{iz}\qty(\hat{z} \cross \vectorbold{B}^*)_{j\alpha}\big)\Big)\nonumber\\
&&\quads[2]{}+\frac{\gamma^4}{\qty(\gamma+ 1)^2}\beta^4\delta_{iz} \delta_{jz} E_{z\alpha} E^*_{z\alpha}\Big].\nonumber
\end{eqnarray}

In calculating the $E^{\,\prime}_{zz}$ component, where we recognize that $(\hat{z} \cross \vectorbold{B})$ vanishes, and get
\begin{eqnarray}
    &&\expval*{\hat{E}^{\,\prime \,2}_z(z,t)}_0=\nonumber\\
    &&\quads[2]\qty(\gamma^2-2\frac{\gamma^3}{\gamma+ 1}\beta^2+\frac{\gamma^4}{\qty(\gamma+ 1)^2}\beta^4)\sum_{\alpha} E_{z\alpha} E^*_{z\alpha}\nonumber\\
    &&\quads[2]=\sum_{\alpha} E_{z\alpha} E^*_{z\alpha}=\expval*{\hat{E}^{2}_z(z,t)}_0.
\end{eqnarray}
We see that the $E^{\,\prime}_{zz}$ component of the electric field correlator remains unchanged as a result of the boost. This should not be surprising, as the component of any electric and magnetic field in the direction of the Lorentz boost remain unchanged (Here ${\vectorbold{E}_z'}=\vectorbold{E}_z$ and ${\vectorbold{B}_z'}=\vectorbold{B}_z$).

For the calculation of the two other non-vanishing components of the electric field correlator, we recognize that the second and third term in Eq. \eqref{general E} will vanish for both the $E^{\,\prime}_{xx}$ and $E^{\,\prime}_{yy}$ components. The $E^{\,\prime}_{xx}$ correlator now becomes 
\begin{eqnarray}
    &&\expval*{\hat{E}^{\,\prime \,2}_x(z,t)}_0=\gamma^2\qty(\sum_{\alpha} E_{x\alpha}E^{*}_{x\alpha}+\beta^2\sum_{\alpha}B_{y\alpha}B^*_{y\alpha})\\
    &&\quads[2.5]=\gamma^2\qty(\frac{\pi}{a})^4\frac{2}{3\pi}\qty[-\frac{1}{120}+F(\xi)-\beta^2\qty(\frac{1}{120}+F(\xi))],\nonumber
\end{eqnarray}
\noindent where the terms $\qty(\hat{z} \cross \vectorbold{B})_{i\alpha}E^{*}_{j\alpha}$ and $E_{i\alpha}(\hat{z} \cross \vectorbold{B}^*)_{j\alpha}$ vanish since $\expval*{\hat{E}_i(\vectorbold{r},t) \hat{B}_j(\vectorbold{r},t)}_0=0$. The $E^{\,\prime}_{yy}$ correlator can be computed in the same manner,
\begin{eqnarray}
    &&\expval*{\hat{E}^{\,\prime \,2}_y(z,t)}_0=\gamma^2\qty(\sum_{\alpha} E_{y\alpha}E^{*}_{y\alpha}+\beta^2\sum_{\alpha}B_{x\alpha}B^*_{x\alpha})\\
    &&\quads[2.5]=\gamma^2\qty(\frac{\pi}{a})^4\frac{2}{3\pi}\qty[-\frac{1}{120}+F(\xi)-\beta^2\qty(\frac{1}{120}+F(\xi))].\nonumber
\end{eqnarray}

\noindent It is straightforward to show that the off-diagonal components for the electric field correlator still vanish under the boost.

With the general Lorentz transform of $\vectorbold{B}$, 
\begin{equation}\label{B'}
    \vectorbold{B}\xmapsto{\,\,\,\,\Lambda\,\,\,\,}\vectorbold{B^{\,\prime}}=\gamma\qty(\vectorbold{B}-\boldsymbol{\beta}\crossproduct\vectorbold{E})-\frac{\gamma^2}{\gamma+1}\boldsymbol{\beta}\qty(\boldsymbol{\beta}\dotproduct\vectorbold{B}),
\end{equation}
we see that the magnetic field correlator transforms in the following way
\begin{eqnarray}
    &&\expval*{\hat{B}^{\,\prime}_i(\vectorbold{r},t) \hat{B}^{\,\prime}_j(\vectorbold{r},t)}_0=\sum_{\alpha} B^{\,\prime}_{i\alpha}(\vectorbold{r})B^{\,\prime*}_{j\alpha}(\vectorbold{r})=\nonumber\\
&&{}\sum_{\alpha}\Big[\gamma^2\Big( B_{i\alpha}B_{j\alpha}^{*}+\qty(\boldsymbol{\beta} \cross \vectorbold{E})_{i\alpha}\qty(\boldsymbol{\beta} \crossproduct \vectorbold{E}^*)_{j\alpha}\nonumber\\
&&\quad\quad\quad\quad{}- B_{i\alpha}(\boldsymbol{\beta} \cross \vectorbold{E}^*)_{j\alpha}-\qty(\boldsymbol{\beta} \cross \vectorbold{E})_{i\alpha}B^{*}_{j\alpha}\Big)\\
&&\quad\quad{}-{}\frac{\gamma^3}{\gamma+ 1}\Big(\big(B_{i\alpha}\beta_{j}-\qty(\boldsymbol{\beta} \cross \vectorbold{E})_{i\alpha}\beta_{j}\big)\qty(\boldsymbol{\beta} \dotproduct \vectorbold{B}^*_{\alpha})\nonumber\\
&&\quad\quad\quad\quad\quad\quad{}+\qty(\boldsymbol{\beta} \dotproduct \vectorbold{B}_{\alpha})\big(\beta_{i}B^*_{j\alpha}-\beta_{i}\qty(\boldsymbol{\beta} \cross \vectorbold{E}^*)_{j\alpha}\big)\Big)\nonumber\\
&&\quad\quad{}+{}\frac{\gamma^4}{\qty(\gamma+ 1)^2}\beta_i\beta_j\qty(\boldsymbol{\beta} \dotproduct \vectorbold{B}_{\alpha})\qty(\boldsymbol{\beta} \dotproduct \vectorbold{B}^*_{\alpha})\Big].\nonumber
\end{eqnarray}

\noindent With the boost aligned as before, in the $\hat{z}$-direction, we see that this becomes
\begin{eqnarray}\label{general B}
&&\expval*{\hat{B}^{\,\prime}_i(\vectorbold{r},t) \hat{B}^{\,\prime}_j(\vectorbold{r},t)}_0=\nonumber\\[5pt]
&&\sum_{\alpha}\Big[\gamma^2\Big( B_{i\alpha}B_{j\alpha}^{*}+\beta\pig(\beta\qty(\hat{z}  \cross \vectorbold{E})_{i\alpha}\qty(\hat{z}  \crossproduct \vectorbold{E}^*)_{j\alpha}\nonumber\\
&&\quads[4]{}- B_{i\alpha}(\hat{z}  \cross \vectorbold{E}^*)_{j\alpha}-\qty(\hat{z}  \cross \vectorbold{E})_{i\alpha}B^{*}_{j\alpha}\pig)\Big)\\
&&\quads[2]{}-\frac{\gamma^3}{\gamma+ 1}\beta^2\Big(\big(B_{i\alpha}\delta_{jz}-\qty(\hat{z} \cross \vectorbold{E})_{i\alpha}\delta_{jz}\big) B^*_{z\alpha}\nonumber\\
&&\quads[7]{}+ B_{z\alpha}\big(\delta_{iz}B^*_{j\alpha}-\delta_{iz}\qty(\hat{z} \cross \vectorbold{E}^*)_{j\alpha}\big)\Big)\nonumber\\
&&\quads[2]{}+\frac{\gamma^4}{\qty(\gamma+ 1)^2}\beta^4\delta_{iz} \delta_{jz} B_{z\alpha} B^*_{z\alpha}\Big].\nonumber
\end{eqnarray}

In calculating the $B^{\,\prime}_{zz}$ component, where we again recognize that $(\hat{z} \cross \vectorbold{E})$ vanishes, and get
\begin{eqnarray}
    &&\expval*{\hat{B}^{\,\prime \,2}_z(z,t)}_0=\nonumber\\
    &&\quads[2]\qty(\gamma^2-2\frac{\gamma^3}{\gamma+ 1}\beta^2+\frac{\gamma^4}{\qty(\gamma+ 1)^2}\beta^4)\sum_{\alpha} B_{z\alpha} B^*_{z\alpha}\nonumber\\
    &&\quads[2]=\sum_{\alpha} B_{z\alpha} B^*_{z\alpha}=\expval*{\hat{B}^{2}_z(z,t)}_0.
\end{eqnarray}
The magnetic field correlator aligned with the boost direction remains unchanged, just as we saw before with the electric field component. 

The second and third terms in Eq. \eqref{general B} will vanish for both the $B^{\,\prime}_{xx}$ and $B^{\,\prime}_{yy}$ components. The $B^{\,\prime}_{xx}$ correlator is now
\begin{eqnarray}
    &&\expval*{\hat{B}^{\,\prime \,2}_x(z,t)}_0=\gamma^2\qty(\sum_{\alpha} B_{x\alpha}B^{*}_{x\alpha}+\beta^2\sum_{\alpha}E_{y\alpha}E^*_{y\alpha})\\
    &&\quads[2.5]=\gamma^2\qty(\frac{\pi}{a})^4\frac{2}{3\pi}\qty[-\frac{1}{120}-F(\xi)-\beta^2\qty(\frac{1}{120}-F(\xi))],\nonumber
\end{eqnarray}
\noindent where the terms $\qty(\hat{z} \cross E)_{i\alpha}B^{*}_{j\alpha}$ and $B_{i\alpha}(\hat{z} \cross E^*)_{j\alpha}$ vanish since $\expval*{\hat{E}_i(\vectorbold{r},t) \hat{B}_j(\vectorbold{r},t)}_0=0$. The $B^{\,\prime}_{yy}$ correlator can be computed in the same manner,
\begin{eqnarray}
    &&\expval*{\hat{B}^{\,\prime \,2}_y(z,t)}_0=\gamma^2\qty(\sum_{\alpha} B_{y\alpha}B^{*}_{y\alpha}+\beta^2\sum_{\alpha}E_{x\alpha}E^*_{x\alpha})\\
    &&\quads[2.5]=\gamma^2\qty(\frac{\pi}{a})^4\frac{2}{3\pi}\qty[-\frac{1}{120}-F(\xi)-\beta^2\qty(\frac{1}{120}-F(\xi))].\nonumber
\end{eqnarray}
It is again straightforward to show that the off-diagonal components for the magnetic field correlator still vanish under the boost.

With the newly calculated electric and magnetic field correlators in the boosted frame, we can show how the stress-energy tensor transforms as a result of the transformation of its constituent quantities. We begin with the Lorentz transformation of the quantum version of the Maxwell stress tensor in Eq. \eqref{Maxwell quantum}, which becomes
\begin{eqnarray}
    &&4\pi \expval*{\hat{T}^{\,\prime}_{ij}}_0=\\
    &&\quads[2]\expval*{\hat{E}^{\,\prime}_i\hat{E}^{\,\prime}_j}_0-\frac{1}{2}\delta_{ij}\expval*{\hat{\vectorbold{E}}^{\,\prime\,2}}_0+\expval*{\hat{B}^{\,\prime}_i\hat{B}^{\,\prime}_j}_0-\frac{1}{2}\delta_{ij}\expval*{\hat{\vectorbold{B}}^{\,\prime\,2}}_0.\nonumber
\end{eqnarray}
The $T^{\,\prime}_{zz}$ component is now
\begin{equation}
    \expval*{\hat{T}^{\,\prime}_{zz}}_0= \frac{1}{8 \pi}\bqty{\expval*{\hat{E}^{\,\prime\,2}_{z}}_0-\expval*{\hat{E}^{\,\prime\,2}_{\parallel}}_0+\expval*{\hat{B}^{\,\prime\,2}_{z}}_0-\expval*{\hat{B}^{\,\prime\,2}_{\parallel}}_0}, 
\end{equation}
which simplifies to
\begin{equation}\label{Tzz prime}
\begin{split}
    \expval*{\hat{T}^{\,\prime}_{zz}}_0&=\gamma^2\qty(3+\beta^2)\qty[\frac{\pi^2}{720a^4}].
\end{split}
\end{equation}
It is clear that we recover the zero velocity rest-frame value seen in Eq. \eqref{OG T components}.
In the same way,
\begin{eqnarray}\label{Txx/Tyy prime}
    \expval*{\hat{T}^{\,\prime}_{xx}}_0=\expval*{\hat{T}^{\,\prime}_{yy}}_0&&=-\frac{1}{8\pi }\qty(\expval*{\hat{E}^{\,\prime\,2}_z(z,t)}_0+\expval*{\hat{B}^{\,\prime\,2}_z(z,t)}_0)\nonumber\\
    &&=-\frac{\pi^2}{720a^4}=\expval*{\hat{T}_{xx}}_0=\expval*{\hat{T}_{yy}}_0.
\end{eqnarray}

We see that the perpendicular components of the stress tensor relative to the boost remain unchanged. This is an expected result, as the regularized stress-energy tensor $\Theta^{\mu\nu}$, where $\Theta^{ij}(z)=-T_{ij}(z)$, is invariant with respect to an arbitrary Lorentz boost parallel to the plates \cite{plunien1986casimir}. As in the rest-frame setup, the off-diagonal boosted terms of $\expval*{\hat{T}^{\,\prime}_{ij}}$ vanish. This is an expected result as the infinite parallel plate Casimir setup should not experience additional shear stress when boosted into the new frame.

The energy density, whose rest-frame form is presented in Eq. \eqref{energy density eq}, with the specific value given by Eq. \eqref{energy density value}, can be calculate in the boosted frame as
\begin{eqnarray}\label{u prime}
    u^{\,\prime}(\vectorbold{r},t)=\expval*{\hat{\Theta}^{\,\prime\,00}}^{ren}_0&&{}=\frac{1}{8\pi }\qty(\expval*{\vectorbold{\hat{E}}^{\,\prime\,2}(\vectorbold{r},t)}_0+\expval*{\vectorbold{\hat{B}}^{\,\prime\,2}(\vectorbold{r},t)}_0)\nonumber\\
    &&{}=-\gamma^2(1+3\beta^2)\qty[\frac{\pi^2}{720 a^4}].
\end{eqnarray}
It is now necessary to calculate the momentum density terms of the stress-energy tensor, $\expval*{\hat{\Theta}^{0j}}^{ren}_0$, in the boosted frame. With the definition
\begin{eqnarray}\label{momentum gen comp}
    \expval*{\hat{\Theta}^{\,\prime\,0j}}^{ren}_0=\frac{1}{4\pi}\qty(\expval*{\vectorbold{\hat{E}^{\,\prime}}\crossproduct\vectorbold{\hat{B}^{\,\prime}}}_0)_j,
\end{eqnarray}
it is clear to see that in the rest-frame of the Casimir cavity there is no net momentum since $\expval*{\hat{E}_i(\vectorbold{r},t) \hat{B}_j(\vectorbold{r},t)}_0=0$. This will not be the case in the boosted frame. Here, the subscript $j$ denoting the momentum direction is replaced by $k$ to make the cross product notation more agreeable. With this change, the right-hand side of Eq. \eqref{momentum gen comp} will be expressed as
\begin{equation}\label{levi civita form}
    \begin{split}
    \frac{1}{4\pi}\qty(\expval*{\vectorbold{\hat{E}^{\,\prime}}\crossproduct\vectorbold{\hat{B}^{\,\prime}}}_0)_k=\frac{1}{4\pi}\Bigl(\epsilon^{ijk}\sum_\alpha\hat{E}^{\,\prime}_{i\alpha}\hat{B}^{\,\prime*}_{j\alpha}\Bigr)_k.
    \end{split}
\end{equation}
The mixed boosted correlator term can be expanded out with the transformed electric and magnetic field terms in Eqs. \eqref{E'} and \eqref{B'}, respectively. This now becomes
\begin{eqnarray}
    &&\sum_\alpha\hat{E}^{\,\prime}_{i\alpha}\hat{B}^{\,\prime*}_{j\alpha}=\nonumber\\
    &&\quad\sum_{\alpha}\Big[\gamma^2\Big(E_{i\alpha}B_{j\alpha}^{*}-\qty(\boldsymbol{\beta} \cross \vectorbold{B})_{i\alpha}\qty(\boldsymbol{\beta} \crossproduct \vectorbold{E}^*)_{j\alpha}\nonumber\\
    &&\quads[5]{}- E_{i\alpha}(\boldsymbol{\beta} \cross \vectorbold{E}^*)_{j\alpha}+\qty(\boldsymbol{\beta} \cross \vectorbold{B})_{i\alpha}B^{*}_{j\alpha}\Big)\\
    &&\quads[3]{}-\frac{\gamma^3}{\gamma+ 1}\Big(\big(E_{i\alpha}\beta_{j}+\qty(\boldsymbol{\beta} \cross \vectorbold{B})_{i\alpha}\beta_{j}\big)\qty(\boldsymbol{\beta} \dotproduct \vectorbold{B}^*_{\alpha})\nonumber\\
    &&\quads[7]{}+\qty(\boldsymbol{\beta} \dotproduct \vectorbold{E}_{\alpha})\big(\beta_{i}B^*_{j\alpha}-\beta_{i}\qty(\boldsymbol{\beta} \cross \vectorbold{E}^*)_{j\alpha}\big)\Big)\nonumber\\
    &&\quads[3]{}+\frac{\gamma^4}{\qty(\gamma+ 1)^2}\beta_i\beta_j\qty(\boldsymbol{\beta} \dotproduct \vectorbold{E}_{\alpha})\qty(\boldsymbol{\beta} \dotproduct \vectorbold{B}^*_{\alpha})\Big].\nonumber
\end{eqnarray}
Any mixed terms containing both $E_{i\alpha}$ and $B^{*}_{j\alpha}$, along with their complex conjugates, will vanish. Additionally, once the boost is aligned in the $\hat{z}$-direction, any terms formed from a mixing of $(\hat{z}\dotproduct\vectorbold{A})$ and $(\hat{z}\crossproduct\vectorbold{A})_{i\alpha}$, with appropriate conjugates in place, will vanish for both $\vectorbold{A}=\vectorbold{E,B}$. This leaves us with
\begin{eqnarray}
        \sum_\alpha\hat{E}^{\,\prime}_{i\alpha}\hat{B}^{\,\prime*}_{j\alpha}&&{}=\gamma^2\beta\sum_{\alpha}\qty[\qty(\hat{z} \cross \vectorbold{B})_{i\alpha}B^{*}_{j\alpha}- E_{i\alpha}(\hat{z} \cross \vectorbold{E}^*)_{j\alpha}]\nonumber\\
        &&{}=\gamma^2\beta\sum_{\alpha}\Big[\qty(-B_{y\alpha}\delta_{ix}+B_{x\alpha}\delta_{iy})B^{*}_{j\alpha}\nonumber\\
        &&\quads[5]{}+ E_{i\alpha}\qty(E^{*}_{y\alpha}\delta_{jx}-E^{*}_{x\alpha}\delta_{jy})\Big].
\end{eqnarray}

We are finally left with
\begin{eqnarray}
     &&\sum_\alpha\hat{E}^{\,\prime}_{i\alpha}\hat{B}^{\,\prime*}_{j\alpha} =\\
    &&\begin{dcases*}
    -\gamma^2\beta\sum_{\alpha}\qty(B_{y\alpha}B^{*}_{y\alpha} +E_{x\alpha}E^{*}_{x\alpha} )&
      for $i=x$ and $j=y$,\\
    \gamma^2\beta\sum_{\alpha}\qty(B_{x\alpha}B^{*}_{x\alpha} +E_{y\alpha}E^{*}_{y\alpha})  &
      for $i=y$ and $j=x$,\\
    \omit\hfil$0$\hfil & otherwise.
    \end{dcases*}\nonumber
\end{eqnarray}
Substituting this into Eq. \eqref{levi civita form}, and contracting against the Levi-Civita symbol, we see that the only non-vanishing momentum component is along the boosted $\hat{z}$-direction, as expected, and equal to 
\begin{eqnarray}\label{momentum prime}
    \expval*{\hat{\Theta}^{\,\prime\,0z}}^{ren}_0&&{}=-\frac{\gamma^2\beta}{4\pi}\Big(\expval*{\hat{B}^{2}_x(z,t)}_0+\expval*{\hat{B}^{2}_y(z,t)}_0\nonumber\\
    &&\quads[4]+\expval*{\hat{E}^{2}_x(z,t)}_0+\expval*{\hat{E}^{2}_y(z,t)}_0\Big)\nonumber\\
        &&{}=4\gamma^2\beta\qty[\frac{\pi^2}{720a^4}].
\end{eqnarray}

With this, we now have all the necessary components to verify that a Lorentz transformation of the underlying fields that construct the quantum correlators used to calculated the parallel plate Casimir cavity is exactly equal to the Lorentz transformation of the renormalized stress-energy tensor. Applying the relation $\Theta^{ij}(z)=-T_{ij}(z)$ to the transformed quantum Maxwell stress tensor components in Eqs. \eqref{Tzz prime} and \eqref{Txx/Tyy prime}, along with the directly computed energy and momentum terms of the transformed stress-energy tensor in Eqs. \eqref{u prime} and \eqref{momentum prime}, we arrive at the full expression for the boosted stress-energy tensor in Eq. \eqref{Boosted Cavity}.

%%%%%%%%%%%%%%%%%%%%%%%%%%%%%
%%%%%%%%%%%%%%%%%%%%%%%%%%%%%

\textit{Conclusion}.---In this paper, we have shown that a Lorentz boost of the underlying electromagnetic field correlators of a parallel plate Casimir cavity will generate a renormalized stress-energy tensor which agrees with the direct calculation of the Lorentz boost of the rest-frame stress-energy tensor. This transformation was taken in the direction perpendicular to the surface of the plates, as the system is unaffected by an arbitrary Lorentz boost parallel to the plates. In the rest-frame, the electromagnetic field correlators possess divergences, but the resulting stress-energy tensor will be finite as these infinities exactly cancel. We find that the same applies to the boosted frame, whose correlators continue to possess divergences but transformed such that the resulting stress-energy tensor components retained the correct finite values. 

While this derivation is not novel per se, we do hope that its explicit form is useful in enabling future calculations of Casimir phenomenon involving moving configurations. Of particular interest would be an an analysis of the effect of this boost on the plate geometry itself which determines if the change in Casimir energy density corresponds to a plate setup with length-contracted plate separation. Additionally, an extension of this work to the finite plate case would be interesting, as in this case boosts in the $x$-direction are no longer trivial since fringing fields would transform to nonidentical fields. 

We additionally note that this method of computed boosted energy densities is generally applicable to any Casimir situation in which one knows rest frame field correlators. Of particular interest would be applying this to situations involving more complex geometries such as skewed plates or spheres. In such cases, we expect it is not possible to determine explicit and analytic forms of the correlators, but since our transform acts linearly, it may be possible to extend this scheme to perturbative or series expansion solutions and pass through the transform at each order. To the best of the authors' knowledge, this is the only method in literature for computing Casimir energies and forces in moving frames using field quantities.

\begin{acknowledgments}
We wish to acknowledge David McNutt and Eric Davis for their helpful discussions and reviews.
\end{acknowledgments}

% The \nocite command causes all entries in a bibliography to be printed out
% whether or not they are actually referenced in the text. This is appropriate
% for the sample file to show the different styles of references, but authors
% most likely will not want to use it.
%\nocite{*}

\bibliographystyle{unsrt-phys-eucos}
\bibliography{ref}% Produces the bibliography via BibTeX.

\end{document}